\documentclass[11pt]{article}
\usepackage[latin1]{inputenc}
\usepackage{latexsym}
\usepackage[pdftex]{graphics}
\usepackage{amsmath, verbatim, epsfig, fullpage, amssymb}
\begin{document}
\large
\renewcommand{\baselinestretch}{1}
\normalsize
\renewcommand{\baselinestretch}{1}
\parindent .5cm

\renewcommand{\baselinestretch}{2}
\normalsize
\renewcommand{\baselinestretch}{2}

\markright{
     Schoenberg: Nonparametric estimation of Hawkes productivities. 
     }

\begin{center} {\bf Nonparametric estimation of variable productivity Hawkes processes}\\[.5in] 
\end{center}

\begin{center}
Frederic Paik Schoenberg$^1$. 
\end{center}

{\bf Abstract.} An extension of the Hawkes model where the productivity is variable is considered. 
In particular, the case is considered where each point may have its own productivity 
and a simple analytic formula is derived for the maximum likelihood estimators of these productivities. 
This estimator is compared with an empirical estimator and ways are explored of stabilizing both estimators by lower truncating, smoothing, and rescaling the estimates. 
Properties of the estimators are explored in simulations, and the methods are applied to seismological and 
epidemic datasets to show and quantify substantial variation in productivity. 

\noindent $^1$Department of Statistics,
University of California, Los Angeles, CA 90095, USA, frederic@stat.ucla.edu. 


\section{Introduction.}
The Hawkes point process model (Hawkes 1971), a type of branching point process model, has been used for a wide variety of applications, including seismology (Ogata 1998), invasive species (Balderama et al.\ 2012), disease epidemics (Meyer et al.\ 2012), reported crimes (Mohler et al.\ 2011), terrorist attacks (Clauset and Woodard 2013), and financial events (Bacry et al.\ 2015). 
Hawkes models have long been used in seismology to describe the rate of aftershock activity following an earthquake (Ogata 1988, Ogata 1998) and have outperformed alternatives for earthquake forecasting (Zechar et al.\ 2013, Gordon et al.\ 2015).\\ 

The Hawkes model, in its simplest and original form (Hawkes 1971), is given by 
\begin{eqnarray} 
\lambda(t) = \mu + \int \limits _0 ^t K g(t-u) dN(u), 
\end{eqnarray} 
and has since been extended to the spatial-temporal and marked point process cases. 
Such a model was called {\sl epidemic} by Ogata (1988), since it posits that an 
earthquake can produce aftershocks which in turn produce their own aftershocks, etc.
The parameter $\mu$ governs the background rate, the parameter $K$, called the productivity, 
indicates the expected number of first 
generation offspring are triggered by a given point, and the triggering density g dictates how far the offspring points are from the points triggering them. 
In the spatial-temporal case, $\mu$ and $g$ are in general functions of both space and time, and $g$ may also depend on the marks of prior points or on external covariates if this information is available (Ogata 1998).\\ 

The focus of this paper is on the productivity, $K$. 
In previous extensions of the Hawkes process, K has been allowed to vary, rather than be constant for all points. 
In particular, for the Epidemic-Type Aftershock Sequence (ETAS) model commonly used to describe earthquake occurrences, $K$ is an exponential function of the magnitude of the triggering earthquake, based on the observation that larger earthquakes 
tend to have exponentially more aftershocks (Ogata 1988).  
Harte (2014), Wetzler et al.\ (2016), and Dascher-Cousineau et al.\ (2020) also allow the productivity to vary for different earthquakes. 
In an application to disease epidemics, Schoenberg et al.\ (2019) allowed the productivity to be a power-law function of the conditional intensity at the point in question.\\

Here, we allow each point to have its own productivity and consider the simultaneous estimation of all these productivities, without any parametric constraints on how the productivity varies over time. 
If $n = N(0,T)$ points are observed between time $0$ and time $T$, there are thus $n$ productivities to estimate, and we consider estimating them by maximum likelihood. 
The resulting estimates will then have very large variance but may be smoothed to produce more stable estimates.\\ 


The structure of this paper is as follows.
Following a brief review of some mathematical preliminaries and a specification of the variable productivity Hawkes model in Section 2, 
we derive an analytic solution for the maximum likelihood estimates and empirical estimates of the productivities for the variable productivity Hawkes process in Section 3. 
Section 4 discusses smoothing and other ways to improve the stability of these estimators, and the performance of the estimators is explored via simulations in Section 5. 
Section 6 applies the estimation procedure to a catalog of earthquakes and an epidemic dataset, and 
concluding remarks are given in Section 7.\\ 

\section{Variable productivity Hawkes models. }

A point process is a collection of points $\{\tau_1, \tau_2, ...\}$ 
occurring in some metric space $S$ (Daley and Vere-Jones, 2003; Daley and Vere-Jones, 2007). 
Frequently in applications the points occur in time, or in space and time. 
Such processes are typically modeled via their conditional rate (also called conditional intensity), 
$\lambda(t | \mathcal{H}_t)$ or $\lambda(s | \mathcal{H}_s)$, which represents the infinitesimal rate at which points are 
accumulating at time $t$ or at location $s$ of space-time, 
given information on all points occurring prior. 
For maximal generality, in what follows we will assume the metric space is a portion of space-time and 
for simplicity we will write the conditional intensity as $\lambda(s)$, suppressing the dependence on the history $\mathcal{H}_s$.\\ 

We consider the variable-productivity Hawkes process model 
\begin{eqnarray} 
\lambda(s) = \mu(s) + \int \limits K(u) g(s-u) dN(u), \label{Ks}
\end{eqnarray} 
where $\mu > 0$, $g \geq 0$ is a triggering density satisfying $g(t',x',y') = 0$ for $t' < 0$, and $\int_0^\infty g(s') ds' = 1$, and in general $s = (t,x,y) \in S$ is a spatial-temporal location, though all of our results will also apply for a purely temporal point process where $s$ is simply time. 
The triggering density $g$
describes the secondary activity induced by a prior event,
and the constant $K$ is the productivity, which is typically required to satisfy $0 \leq K < 1$ in order to ensure
stationarity (Hawkes, 1971). 
Several forms of the triggering function $g$ have been posited for describing seismological data,
such as 
$g(u_i ; m_i)
= \frac{1}{(u_i+c)^{\rho}}\, e^{a(m_i-M_0)}$,
where $u_i=t-\tau_i$ is the time elapsed since event $i$, 
and $M_0$ is the lower cutoff magnitude for the earthquake catalog (Ogata 1988).\\ 

Some special cases are worth considering. 
If $K(\tau_i)$ is a constant, then the model (\ref{Ks}) corresponds to the ordinary Hawkes process. 
If $K(\tau_i) = f(\tau_i)$, for some fixed function $f$, then there is a trend in the productivity, i.e.\ the productivity may increase or decrease as time varies. 
The case $K(\tau_i) = f(m_i)$ includes the ETAS model of Ogata (1988, 1998), where the productivity of earthquake $i$ depends only on its magnitude. 
The case $K(\tau_i) = f\{\lambda(\tau_i)\}$ corresponds to the {\sl recursive} model of Schoenberg et al.\ (2019), where the productivity depends on the conditional intensity, which in turn depends on the productivity of prior points. 
If $K(\tau_i) = f(\tau_i - \tau_{i-1})$, then the process may be called a renewal productivity Hawkes model, as the productivity only depends on the time elapsed since the previous point. 
Another special case is the model of Harte (2014), a generalization of the renewal productivity Hawkes process where the productivity of point $\tau_i$ depends not merely on the single previous point but on whether $\tau_i$ is among a large cluster of prior points.\\ 


Point process models such as Hawkes processes
are typically fit by maximizing the log-likelihood function 
\begin{eqnarray} 
    \ell (\theta) = \sum_{i}
\log\left(\lambda(\tau_i)\right) -
   \int _{S} \lambda(s) ds \label{like}
\end{eqnarray}
where $\theta$ 
is the parameter vector to be estimated 
(Daley and Vere-Jones, 2003). \\

Maximum likelihood estimates (MLEs), i.e.\ values of the parameters optimizing equation (\ref{like}), can be searched for by 
conventional gradient-based methods, or via the somewhat more robust iterative procedure in Veen and Schoenberg (2008) where the estimated branching structure probabilities are incorporated into the procedure. 
Under rather general conditions, MLEs are consistent, asymptotically normal, and efficient (Ogata, 1978), and estimates of their variance can be derived 
from the negative of the diagonal elements of the Hessian of the likelihood function (Ogata 1978, Rathbun and Cressie 1994). These estimated variances can be used to construct estimates of standard errors and $95\%$-confidence bounds.\\ 


\section{Proposed estimators.} 

For the variable productivity Hawkes model (\ref{Ks}), consider the estimation of $K(s)$, 
for $s = \tau_1, \tau_2, ..., \tau_n$. 
For simplicity, assume for the moment that $\mu$ and $g$ are known, and $\mu(s)=\mu$ is a constant. 
When the parameters $\mu$ and $g$ are unknown, 
one option is to estimate them by fitting a simple Hawkes process with constant productivity by maximum likelihood, and using the resulting estimates for the variable productivity Hawkes model.\\ 

Setting the partial derivatives of $\ell$ in equation (\ref{like}) with respect to each value of $K(\tau_i)$ 
to zero, for $i = 1, ..., n-1$,
reduces to  
\begin{eqnarray} \label{lik2}
0 &=& \partial \ell (\theta) / \partial K(\tau_i) \nonumber \\ 
& = & \sum \limits _{j: j>i} \frac{\partial \lambda (\tau_j) / \partial K(\tau_i)}{\lambda(\tau_i)} - \partial / \partial K(\tau_i) [\int \limits _{S} \lambda(s) ds] \nonumber \\ 
& = & \sum \limits _{j=i+1}^{n} \frac{g(\tau_j-\tau_i)}{\lambda(\tau_j)} - 1,
\end{eqnarray}
for each $i = 1, 2, ..., n-1$.\\ 

Note that $\lambda(\tau_1) = \mu$, and relation (\ref{lik2}), for $i = 1, 2, ..., n-1,$ can be viewed as $n-1$ linear equations in the $n-1$ unknowns, $\{1/\lambda(\tau_2), 1/ \lambda(\tau_3), ..., 1/ \lambda(\tau_n)\}$. 
One can thus readily solve for $1/\lambda(\tau_i)$ using these equations, and consequently obtain estimates of 
$\lambda(\tau_i)$, for $i = 2, 3, ..., n$. 
Further, evaluating the conditional rate in equation (\ref{Ks}) at the first $n-1$ observed points $\tau_1, ..., \tau_{n-1}$, one obtains the $n-1$ equations 
\begin{eqnarray} \label{secondeqns}
\lambda(\tau_j) = \mu + K(\tau_i) \sum \limits _{i<j} g(\tau_j - \tau_i), 
\end{eqnarray}
 for $j = 1, 2, ..., n-1$, 
 which are linear in the $n-1$ unknowns $\{K(\tau_1), K(\tau_2), ..., K(\tau_{n-1})\}$. 
 This yields maximum likelihood estimates of $K(\tau_j)$ for $j = 1, ..., n-1$, and the MLE of $K(\tau_n)$ is 0.\\ 

With a bit of additional notation we may write the resulting estimator in a very simple and condensed form as follows.
For any vector $z = \{z_1, z_2, ..., z_k\}$, let $1/z$ represent the vector
$\{1/z_1, 1/z_2, ..., 1/z_k\}$.
Let $G$ denote the $(n-1) \times (n-1)$ upper triangular matrix with entries $G[i,j] = g(\tau_{j+1} - \tau_i)$, for $i \leq j$, and $G[i,j] = 0$ otherwise. 
Let ${\bf \lambda}$ represent the $(n-1)$-vector 
$\{\lambda(\tau_2), \lambda(\tau_3), ..., \lambda(\tau_n)\}$, 
let ${\bf 1}$ denote the $(n-1)$-vector $\{1,1,...,1\}$, 
and let $K$ denote the $(n-1)$-vector $K(\tau_1), K(\tau_2), ..., K(\tau_{n-1})$.\\ 

With this notation, equation (\ref{lik2}) can be rewritten as 
\begin{eqnarray}
G (1/ {\bf \lambda}) = {\bf 1}. 
\label{6b}
\end{eqnarray}
The estimate of $1/ {\bf \lambda}$ satisfying equation (\ref{6b}) is thus $G^{-1} {\bf 1}$, assuming $G$ is invertible.
Similarly, equation (\ref{secondeqns}) may be rewritten 
\begin{eqnarray}
{\bf \lambda} = \mu + G^T K , \label{5b}
\end{eqnarray}
whose solution is $\hat K = (G^T)^{-1}({\bf \lambda} - \mu)$. 
Combining these two formulas, the resulting vector $\hat K = \{\hat K(\tau_1), ..., \hat K(\tau_{n-1})$ of estimates may be written
\begin{eqnarray}
\hat K = (G^T)^{-1} [1/(G^{-1} \bf 1) - \mu], \label{maineq}
\end{eqnarray}
and $\hat K(\tau_n) = 0$.\\ 

The estimates obtained via equation (\ref{maineq}) can be computed simply and rapidly 
provided the matrix $G$ is invertible. 
The speed with which the estimates in (\ref{maineq}) may be obtained enables approximate standard errors 
for these estimates to be constructed by repeated simulation and estimation.
However, because it relies on estimating $n$ parameters based on $n$ observed points, 
the estimator (\ref{maineq}) will have very high variance 
and in practice can benefit greatly from smoothing and other methods to decrease variability, as described in the next Section.\\ 

As an alternative to the estimator based on (\ref{maineq}), one may consider the estimator similar to that used in Wetzler et al.\ (2016), which we will refer to as the {\sl empirical} estimator of the productivities. 
Specifically, for each point $\tau_i$, for $i = 1, 2, ..., n$, one may obtain an estimate of its productivity $K_i$ by fixing some time interval $\Delta$ and simply letting $\hat K_i$ equal the number of points occurring in $(\tau_i, \tau_i + \Delta)$ minus $\Delta \mu$. 
Wetzler et al. (2016) suggest using $\Delta = 7$ days.\\ 

\section{Improving stability.} 

The productivity $K$ is typically constrained to be non-negative in order to ensure the proint process (1) is well defined. This suggests improving the stability of the estimator (\ref{maineq}) by lower-truncating the estimates at 0, 
i.e. replacing any estimate $\hat K_i$ with $\max\{\hat K_i, 0\}$.\\ 

In addition, using the martingale formula (see e.g.\ Daley and Vere-Jones, 2003), and the fact that $g$ is a density, 
\begin{eqnarray} 
E(n) & = & E \int \limits _0 ^T dN  \nonumber \\
& = & E \int \limits _0 ^T \lambda(t) dt \nonumber \\
& = & \mu T + E \int \limits _0 ^T \int \limits _0 ^t K(u) g(t-u) dN(u)dt \nonumber \\ 
& = & \mu T + E \int \limits _0 ^T K(u) \left ( \int \limits _0 ^{T-u} g(t) dt \right ) dN(u) \nonumber \\ 
& \approx & \mu T + E \int \limits _0 ^T K(u) dN(u)\nonumber \\ 
& = & \mu T + E \sum \limits _{i=1}^n K(\tau_i), \label{scale2} 
\end{eqnarray} 
for sufficiently large $T$. 
This suggests rescaling the estimates of $K_i$ by a factor of 
\begin{eqnarray} \label{scaling} 
(n-\mu T)/ \sum \hat K_i.\\ 
\end{eqnarray}  

Further, 
when it is reasonable to assume that $K$ is smooth, the estimates in (\ref{maineq}) may be smoothed to provide more stable estimates of $K$. 
This smoothing may be done over time, nonparametrically, using e.g.\ kernel smoothing or splines, for example. 
When estimating the productivity as a function of some covariate or mark, 
it may be sensible to smooth and scale the productivities in the covariate or mark domain. 
For example, when estimating the productivity $K(m)$ as a function of magnitude on a regular grid of magnitude values, $m_j$, of grid width $\delta m$, equation (\ref{scale2}) implies the expected {\sl mean} productivity should be approximately $1 -\mu T/n$, which suggests rescaling the estimates $\hat K(m)$ so that $\sum f(m) K(m) \delta m = 1 - \mu T/n$.\\ 

The method proposed here may be used as a way of investigating the relationship between productivity and various factors such as magnitude. 
Note that estimation of the productivities proposed here is performed absent any information about the relationship between the productivities $K_i$ and time or magnitude. 
For instance, when considering estimation of productivities for the ETAS model, 
the productivities $K_i$ are estimated assuming no information at all about the magnitudes of the events, and, in both the case of the empirical estimator or estimator (\ref{maineq}), are estimated instead purely based on the seismic activity observed. \\ 

\section{Simulations.}

In this Section, simulations are used as a proof of concept and to explore the performance, especially the stability, of the estimator in (\ref{maineq}), for a number of different types of models. 
Since the focus here is on the estimation of the productivities $K_1, ..., K_n$, we assume that the background rate $\mu$ and the triggering density $g$ are known.\\  

\begin{figure}[h] 
\includegraphics[height=3in,width=6in]{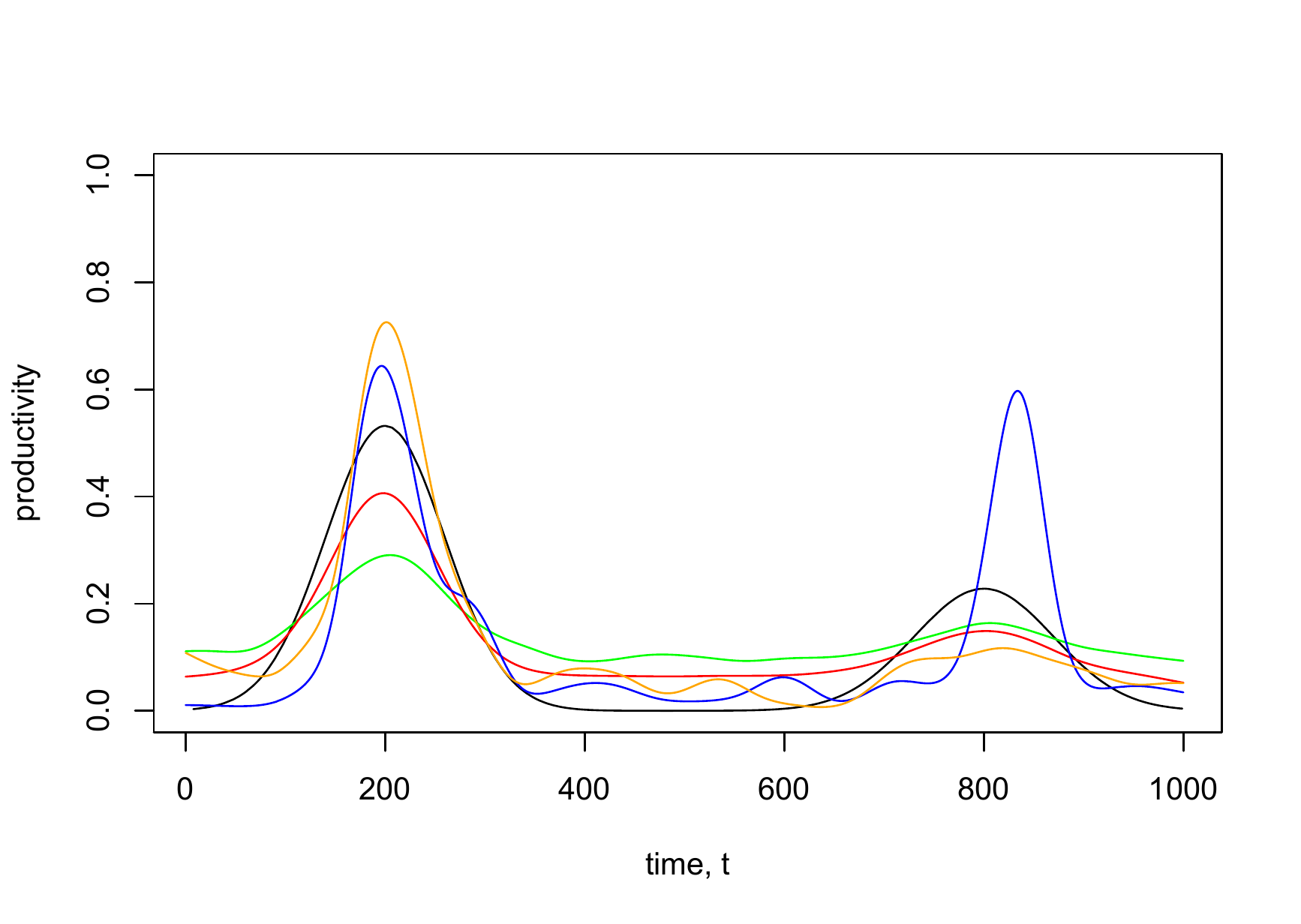} 
\caption{True productivity $K(t) =  80 \phi_1(t) + 40 \phi_2(t)$ (black), where $\phi_1$ and $\phi_2$ are normal densities with means 200 and 800, respectively, and with standard deviations 60 and 70, respectively; estimated productivity using (\ref{maineq}) for one simulated variable-productivity point process (blue); estimated productivity using the empirical estimate for the same simulated process (orange); average of 1000 estimates using (\ref{maineq}) for 1000 simulated point processes (green); average of 1000 empirical estimates for the same 1000 simulated processes (red). The estimates of $K$ were all truncated, scaled using (\ref{scaling}) and smoothed over time using a Gaussian kernel smoother and bandwidth selected using the rule of thumb of Silverman (1986). 
Each simulated variable-productivity point process has exponential triggering function with rate $\beta = 0.7$ and constant background rate $\mu = 0.5$.} 
\label{fig:normalsbrodsky}  
\end{figure} 

The variability in the raw estimator suggested in (\ref{maineq}) is very large. 
Consider, for instance, the case of estimating the productivities of a variable-productivity point process (2) on $[0,1000]$ with 
$\mu = 0.5$, exponential triggering function $g(u) = \beta \exp\{-\beta u\}$ and 
productivity $K(t) =  80 \phi_1(t) + 40 \phi_2(t)$, where $\phi_1$ and $\phi_2$ are normal densities with means 200 and 800, respectively, and with standard deviations 60 and 70, respectively. 
This productivity function is shown in Figure \ref{fig:normalsbrodsky}. 
Using the {\sl raw} estimates of equation (\ref{maineq}), the resulting productivity estimates have an average RMS error of 236.0 for 100 realizations of this process. 
However, the estimates are vastly improved simply by truncating, rescaling and smoothing. 
Indeed, over the 100 realizations, the average RMS error decreased from 4.66 to 0.755 simply by lower truncating the negative productivity estimates at 0 and smoothing the productivities over time using a Gaussian kernel smoother with bandwidth determined using the default of $0.9 \min\{sd, iqr/1.34\} n^{0.2}$
proposed by Silverman (1986), and the average RMS error declined to 0.00874 after truncating, smoothing, and rescaling the smoothed productivity estimates according to (\ref{scaling}).\\ 

Figure \ref{fig:normalsbrodsky} shows the performance of the estimator (\ref{maineq}), after lower truncation at 0, scaling using (\ref{scaling}) and smoothing 
over time using a Gaussian kernel smoother and bandwidth selected using the rule of thumb of Silverman (1986). 
For comparison, the empirical estimator of Wetzler et al. (2016), after also lower-truncating at 0, scaling using (\ref{scaling}), and  smoothing using the same kernel smoother, is shown. 
Both the empirical estimator and the estimator based on (\ref{maineq}) are able to capture the bimodal shape of the productivity function. 
However, the estimator (\ref{maineq}) generally appears to underestimate the height of the peaks in $K$, on average.  
The bandwidth, which was chosen based on considerations in the density estimation context, may be a bit too large to be optimal for estimating this particular productivity function.\\ 

\begin{figure}[h] 
\includegraphics[height=3in,width=6in]{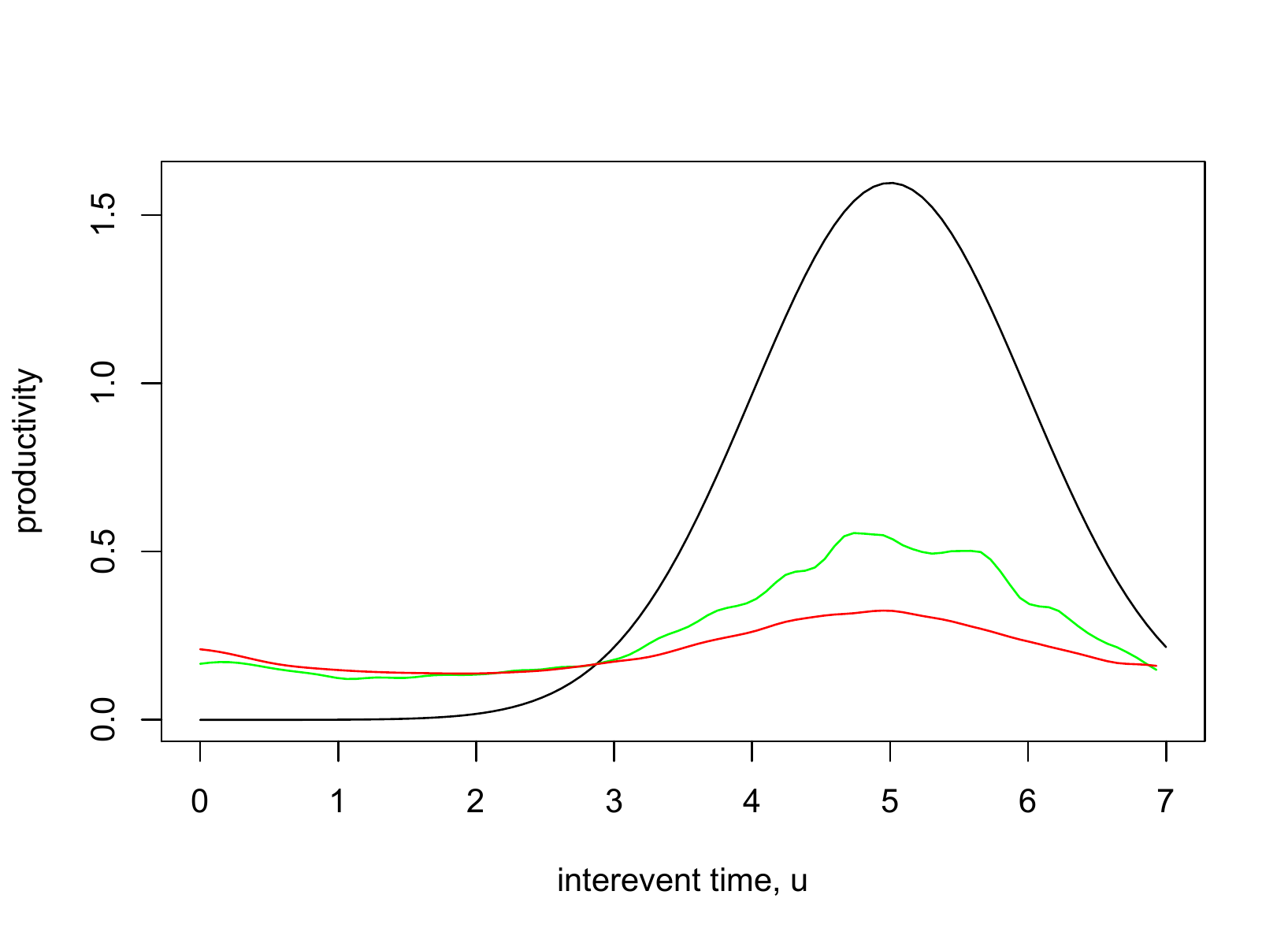} 
\caption{True productivity $K(t_i) =  4 \phi_3(t_i - t_{i-1})$ (black), where $\phi_3$ is the normal densities with means $5.0$ and standard deviation $1.0$, and $t[0] = 0$; mean estimated productivity using (\ref{maineq}) for 1000 simulated point process (green); mean estimated productivity using the empirical estimate for the same simulated processes (red). The estimates of $K$ were all truncated, scaled using (\ref{scaling}) and smoothed over time using a Gaussian kernel smoother and bandwidth selected using the rule of thumb of Silverman (1986). 
Each simulated variable-productivity point process has exponential triggering function with rate $\beta = 0.7$ and constant background rate $\mu = 0.5$. 
}
\label{fig:renewal}  
\end{figure} 

Figure \ref{fig:renewal} shows the performance of the empirical estimator and the estimator from (\ref{maineq}) for estimating a point process with renewal productivity, where the productivity is a function of the time since the previous event. Both estimators are truncated, rescaled, and smoothed 
using a Gaussian kernel smoother and bandwidth selected using the rule of thumb of Silverman (1986). 
The estimators underestimate the peak in the renewal productivity function, though the bias in the 
estimator (\ref{maineq}) is a bit smaller than that of the empirical estimator, after truncating, smoothing and rescaling.\\  

\begin{figure}[h] 
\includegraphics[height=2.8in,width=6in]{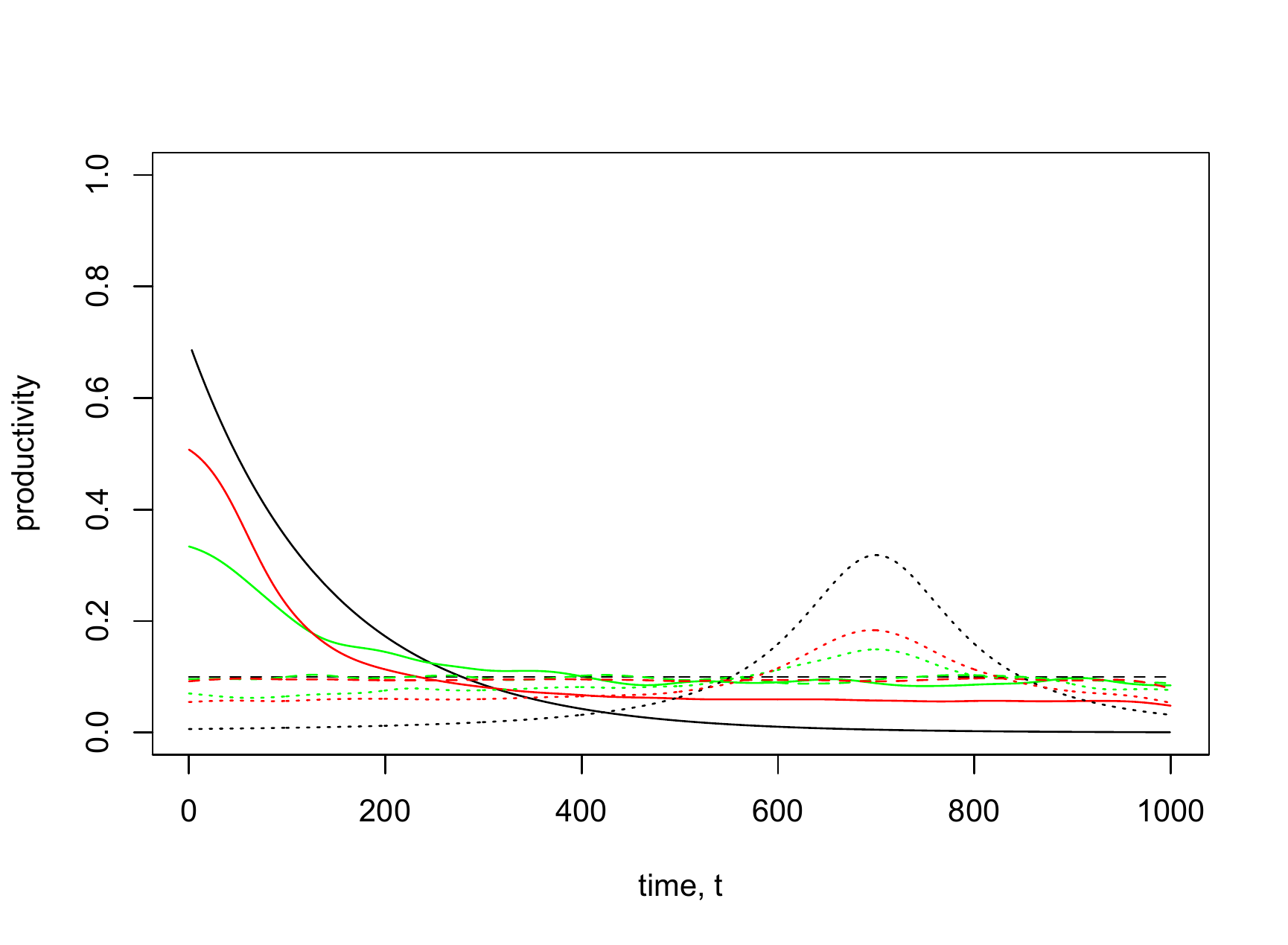}
\caption{True productivity (black curves), estimates using (\ref{maineq}) (green) and empirical estimates (red), after truncating, rescaling and smoothing. Results are averaged over 1000 simulations for each of three different productivity functions. The solid curves correspond to the productivity function $K(t) = 0.7 e^{0.007 t}$, the dashed curves correspond to $K(t) =  0.01$, and the 
dotted curves correspond to $K(t) = 100 \psi(t)$, where $\psi$ is the Cauchy density with location 700 and scale 100. For each simulation, the process was simulated from time $0$ to time $1000$, with exponential triggering function with rate $\beta = 0.7$, and constant background rate $\mu = 0.5$. } 
\label{fig:expadittlewithunifcauchy} 
\end{figure} 

\begin{table} 
\begin{tabular}[h]{|c|c|c|c|} 
\hline 
productivity function & RMSE of unscaled empirical & RMSE of (\ref{maineq}) & RMSE of scaled empirical \\ 
\hline 
normals & 1.75 & 0.187 & 0.0925 \\ 
exponential & 1.90 & 0.171 & 0.0912 \\ 
constant &  1.08 & 0.121 & 0.0570 \\ 
Cauchy & 1.23 & 0.210 & 0.188 \\
renewal & 1.14 & 0.761 & 0.626 \\
\hline 
\end{tabular} 
\caption{RMSE of estimates using (\ref{maineq}) and empirical estimates, averaged over 1000 simulations, after truncating, smoothing, and in the case of (\ref{maineq}) and the scaled empirical estimates, rescaling using (\ref{scaling}). Raw empirical estimates are not rescaled. 
Productivity functions are a mixture of normals where $K(t) = 80 \phi_1(t) + 40 \phi_2(t)$, exponential $K(t) = 0.7 e^{0.007 t}$, constant $K(t) =  0.01$, Cauchy $K(t) = 100 \psi(t)$, and renewal $K(t_i) = 4 \phi_3(t_i - t_{i-1})$, 
where $\phi_1$ is the density of a $N(200,60^2)$ random variable, $\phi_2$ is the $N(800,70^2)$ density, $\phi_3$ is the $N(5,1)$ density, and $\psi$ is the Cauchy density with location 700 and scale 100. For each simulation, the process was simulated from time $0$ to time $1000$, with exponential triggering function with rate $\beta = 0.7$, and constant background rate $\mu = 0.5$.}
\end{table} 

\begin{figure}[h] 
\includegraphics[height=3in,width=6in]{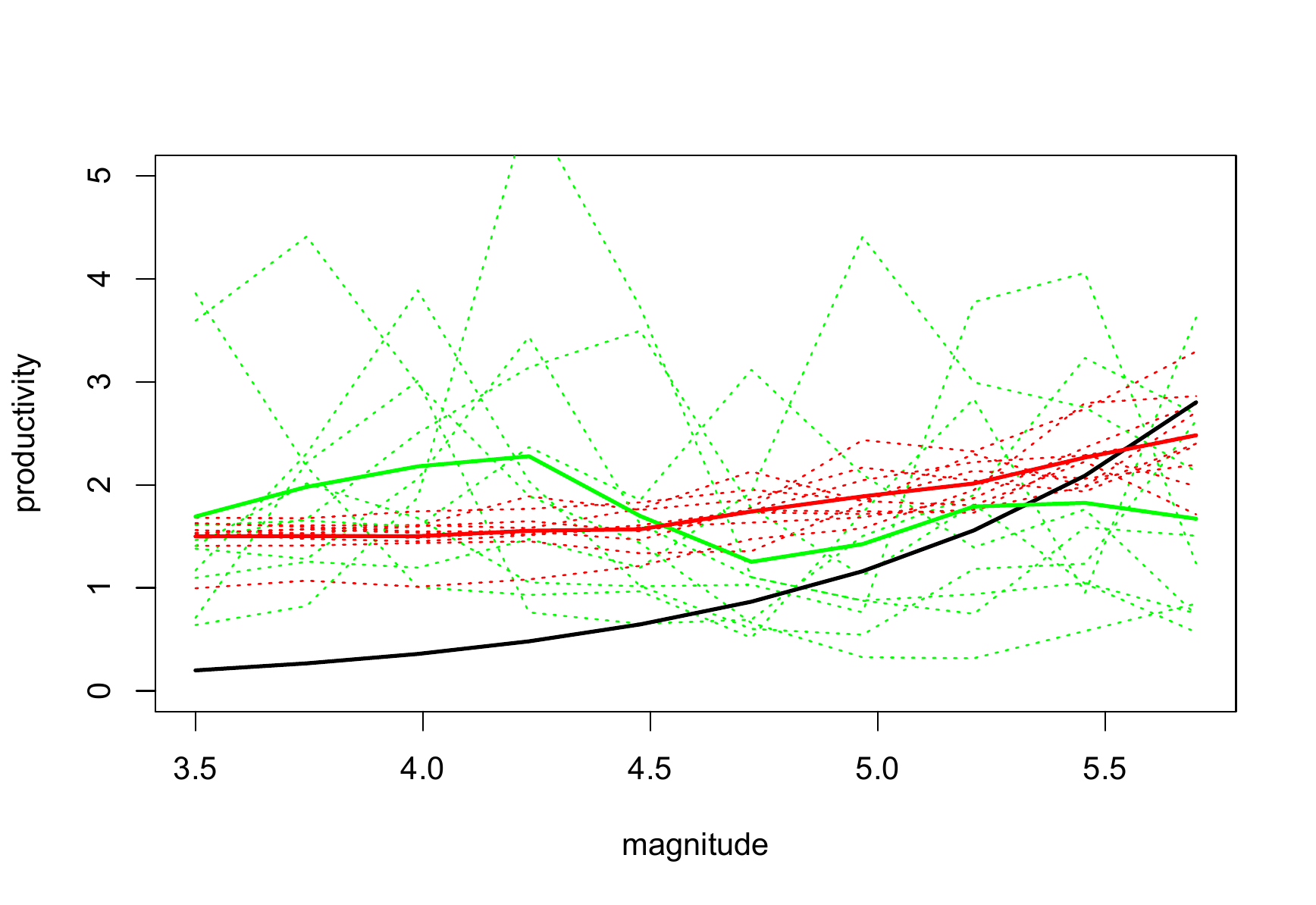}
\caption{True productivity (black curve), estimates using (\ref{maineq}) (green dotted curve) and empirical estimates (red dotted curve), after truncating, rescaling and smoothing, for 10 simulated ETAS processes. 
Solid green and red curves show the average of the 10 estimates for (\ref{maineq}) and the empirical estimates, respectively. Estimates are truncated, rescaled using (\ref{scale2}), and smoothed using a Gaussian kernel smoother with bandwidth obtained using the rule of thumb of Silverman (1986). The simulated model has productivity $K(t) = 0.2 e^{1.2 (m-3.5)}$, simulated from time $0$ to time $1000$, with exponential triggering function with rate $\beta = 2.7$, constant background rate $\mu = 0.1$, and lower-truncated exponential magnitude density with rate 2.3 and truncated at lower magnitude cutoff 3.5. 
The mean RMSEs for these 10 simulations are $1.56$ for the estimates based on (\ref{maineq}) and $0.926$ for the empirical estimates.}  
\label{fig:ETASrescaled} 
\end{figure}

Estimates of several different time-varying productivity functions are shown in Figure \ref{fig:expadittlewithunifcauchy}, and the root mean square errors (RMSEs) of these estimates are reported in Table 1. 
Rescaling, truncating and smoothing result in a very large decrease in the sizes of the errors in every case. 
Figure \ref{fig:ETASrescaled} shows estimates of the productivity function as a function of magnitude for simulated ETAS processes. 
Since for both the empirical estimator and (\ref{maineq}), the estimates are constructed purely based on the temporal patterns in the earthquakes without any use of the earthquake mainshocks, the fact that the estimates are generally able to track the overall shape of the magnitude productivity relationship is surprising. 
However, for ETAS and also for processes where the productivity varies smoothly in time, estimates based on (\ref{maineq}) can be quite unstable. 
The main source of this instability is the fact that (\ref{lik2}) amounts to $n$ equations satisfied by the MLE when simultaneously estimating $n$ parameters $\lambda(\tau_1), ..., \lambda(\tau_n)$, and given $n$ observations, the MLE of $n$ parameters can be highly unstable.\\  

\begin{figure}[h] 
\includegraphics[height=2.8in,width=3in]{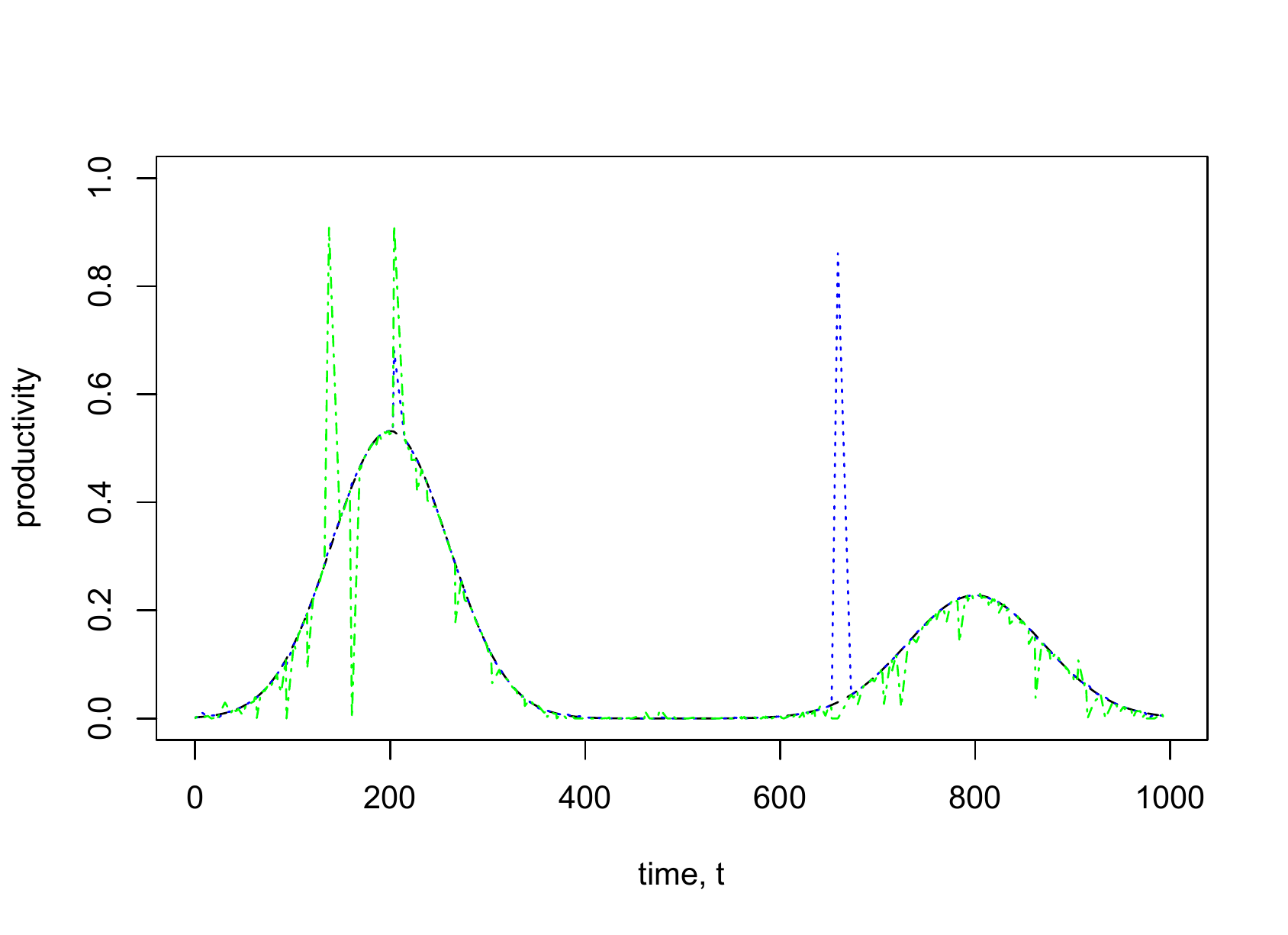}
\includegraphics[height=2.8in,width=3in]{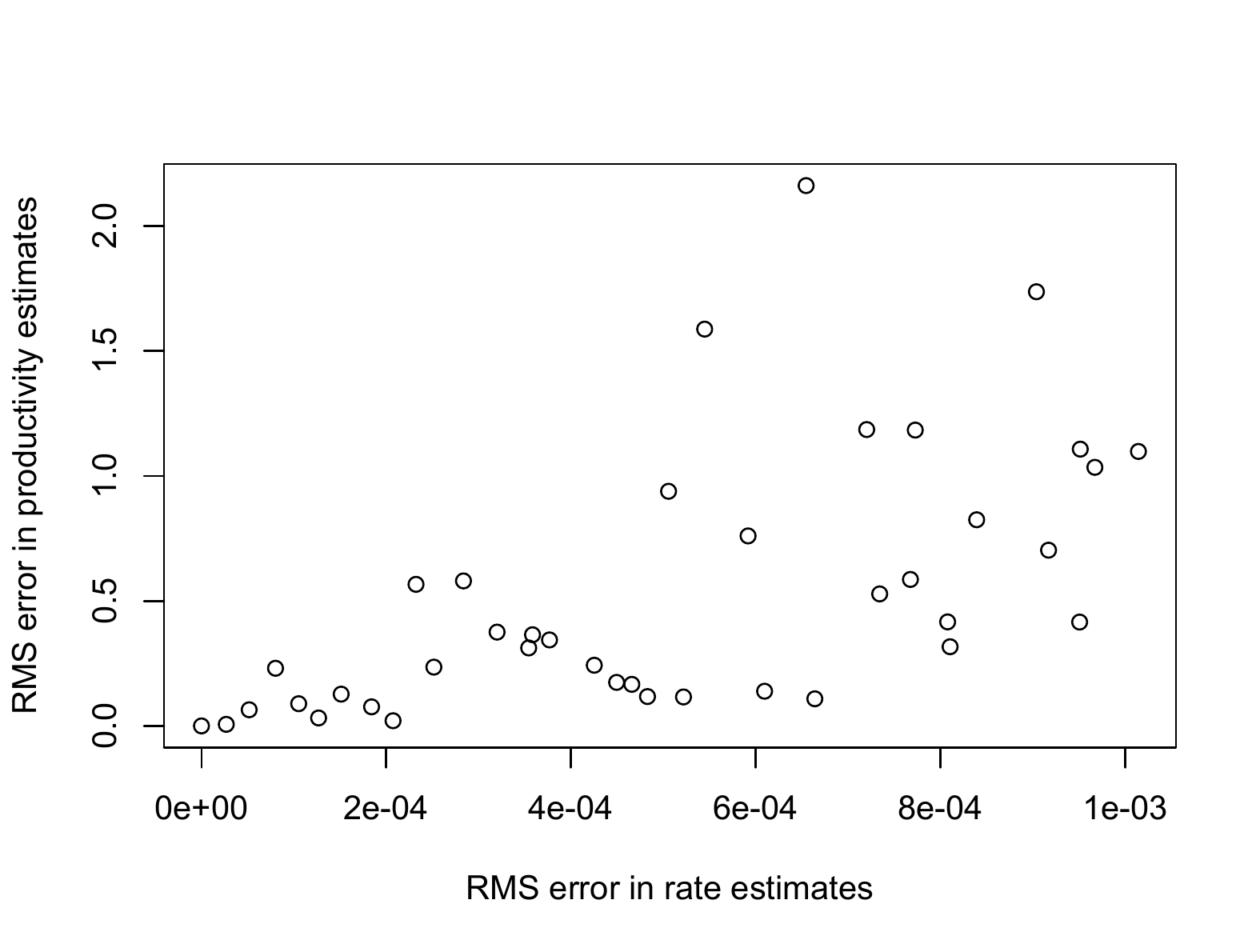} 
\caption{(a) True productivity $K(t) =  80 \phi_1(t) + 40 \phi_2(t)$, indicated by the solid black curve, along with estimated productivities obtained via equation (\ref{secondeqns}) using the true intensities $\lambda(\tau_1),..., \lambda(\tau_n)$ with iid $N(0,\sigma^2)$ noise added to each value of $\lambda(\tau_i)$ for each $i$, where $\sigma = 0.00002$ (green dashed curve) or $\sigma = 0.0005$ (blue dashed curve). 
(b) Root-mean-square (RMS) error in estimates of productivities $K_1, ..., K_n$ using equation (\ref{secondeqns}) as a function of RMS error in estimates of $\lambda(\tau_1), ..., \lambda(\tau_n)$. Each point in the plot corresponds to the same simulated process but with iid $N(0,\sigma^2)$ noise added to each value of $\lambda(\tau_i)$ for each $i$, and where $\sigma^2$ is different for each point  in the plot. The values of $\sigma$ vary from $0$ to $0.001$. The process used has exponential triggering function with rate $\beta = 0.7$, constant background rate $\mu = 0.5$ and productivity $K(t) =  80 \phi_1(t) + 40 \phi_2(t)$, resulting in $567$ points on $[0,1000]$, where $\phi_1$ and $\phi_2$ are normal densities with means 200 and 800, respectively, and with standard deviations 60 and 70, respectively.} 
\label{fig:normalerrorfig} 
\end{figure} 

If the true intensities $\lambda(\tau_1), ..., \lambda(\tau_n)$ were known exactly, rather than estimated, then 
since $\lambda(\tau_1) = \mu$, relation (\ref{secondeqns}) yields $n-1$ linear equations with $n-1$ unknowns, $\hat K(\tau_1), ..., \hat K(\tau_{n-1})$. 
Figure \ref{fig:normalerrorfig} shows how the errors in the productivity estimates $\hat K_1,..., \hat K_n$ would increase with $\sigma$ as small amounts of iid normal noise with mean 0 and variance $\sigma^2$ are added to the true values of $\lambda(\tau_i)$, for $i = 1,..., n$. 
When $\sigma = 0$, the estimates of $K_1, ..., K_{n-1}$ obtained using equation (\ref{secondeqns}) are perfect, and the only source of error in the vector $\hat K$ of productivities is due to the fact that $\hat K_n = 0$. 
As shown in Figure \ref{fig:normalerrorfig}a, when $\sigma = 0.00002$, the estimated productivities are quite accurate, though when $\sigma = 0.0005$, a few of the productivity estimates have substantial errors. 
Indeed, from Figure \ref{fig:normalerrorfig}b, one sees the overall error in the vector $\hat K$ of estimated productivities gradually decreases to very nearly zero as the errors in the estimates of $\lambda$ decrease. 
Thus the bulk of the error when using (\ref{maineq}) is attributable not to equation (\ref{secondeqns}) but to instability in the estimation of the intensity using (\ref{lik2}).\\ 

\section{Applications to Bear Valley seismicity and to Arizona Chlamydia.} 

\begin{figure}[h] 
\includegraphics[height=3in,width=6in]{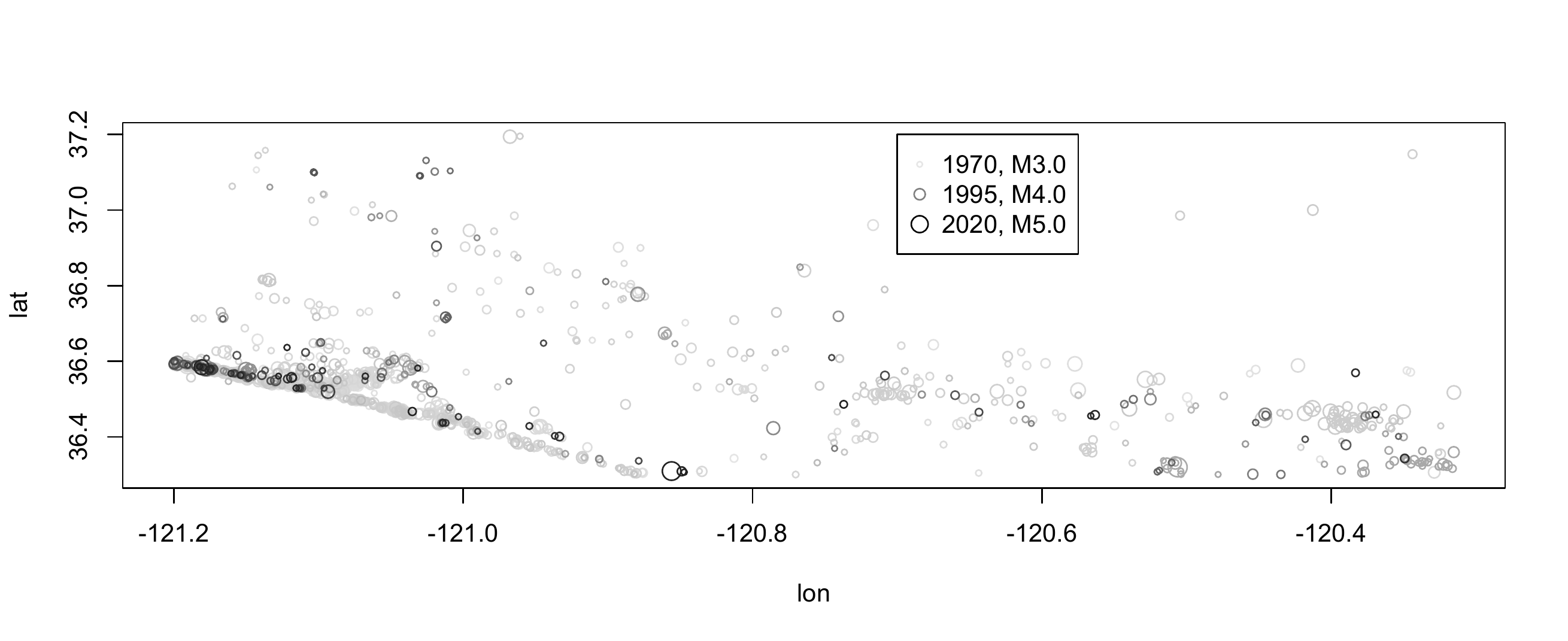}
\caption{Locations, times and magnitudes of recorded Bear Valley earthquakes from 1/1/1970 to 3/6/2020, 
with depth $\leq$ 700km, from the NCEDC dataset. 
More recent events are indicated by darker circles and larger magnitude events correspond to larger circles.} 
\label{fig:bearvalley} 
\end{figure}

We apply our method to analyze the productivities of earthquakes in the Hollister-Bear Valley region, a 35km portion of the San Andreas Fault suggested by Bruce Bolt as an example of seismic hazard calculations and studied in Schoenberg and Bolt (2000) using earthquakes of magnitude at least 3.0 and depth $
\leq$ 700km from 1970-2000.  
Here we include earthquakes from 1/1/1970 to 3/6/2020 and slightly expand the region spatially by $0.2^o$ in each direction to 
latitude 36.3 to 37.2 and longitude -120.3 to -121.2. The depth of the deepest earthquake in the catalog is just 50.84km. 
Data were obtained from the Northern California Earthquake Data Center (NCEDC 2014). 
Figure \ref{fig:bearvalley} shows the epicentral locations of the points along with their 
estimated origin times and magnitudes, with larger and darker points corresponding to higher magnitudes and more recent events, respectively. 
Some of the apparent offset from the San Andreas Fault may be due to errors in location estimates or to epicentral projection for earthquakes at depth.\\  


\begin{figure}[h] 
\includegraphics[height=5in,width=6in]{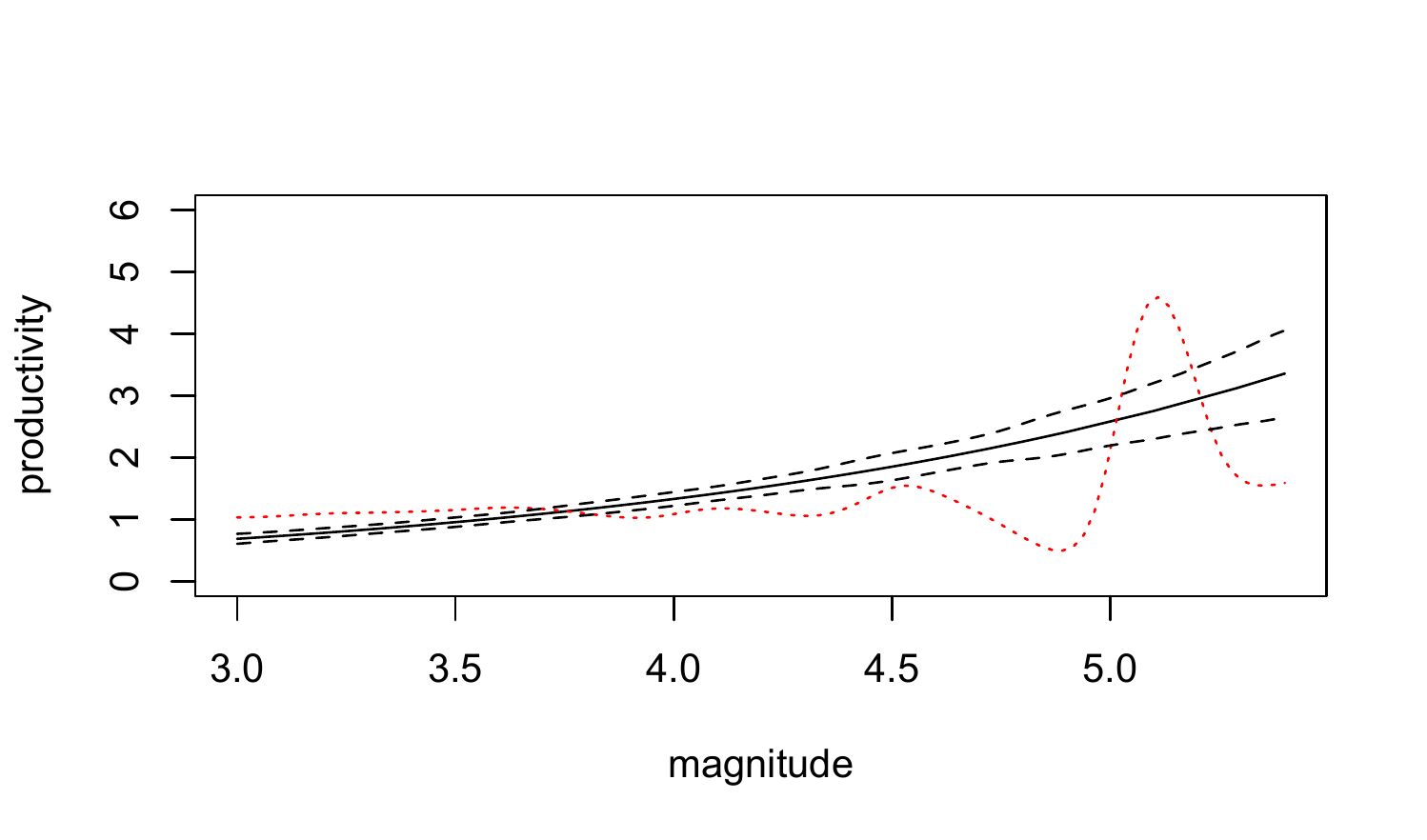}
\caption{Estimated productivity as a function of magnitude for the Bear Valley earthquake data. 
The solid black curve indicates the fitted parametric productivity function when the spatial temporal ETAS model of Ogata (1998) is fit by maximum likelihood. 
The dashed red curve is the truncated, rescaled, and smoothed empirical estimator of productivity as a function of magnitude. 
The black dotted curves show the parametric exponential fitted productivity estimate plus or minus 2 standard errors obtained by simulating 100 ETAS processes with parameters given by the values estimated by MLE to the Bear Valley dataset and re-estimating the parameters for each simulation using the truncated, rescaled, and smoothed empirical estimator.} 
\label{fig:variable} 
\end{figure} 

Figure \ref{fig:variable} shows estimates of the productivity as a function of magnitude for the Bear Valley data. 
The differences between the truncated, rescaled and smoothed empirical estimate and the fitted exponential function in the ETAS model of Ogata (1998) appears to be statistically significant, extending outside the confidence bounds obtained via simulation. 
Here the null hypothesis is that the ETAS model with parameters given by those fit by MLE to this dataset is actually correct, and the dotted curves show the parametric estimates of productivity plus or minus 2 standard errors obtained by simulating 100 ETAS processes and re-estimating the productivity as a function of magnitude for each simulation using the truncated, rescaled and smoothed empirical estimate. 
The results indicate lack of fit of the ETAS model, so the nonparametric productivity estimate may be preferable in this case.\\ 

\begin{figure}[h] 
\includegraphics[height=3in,width=6in]{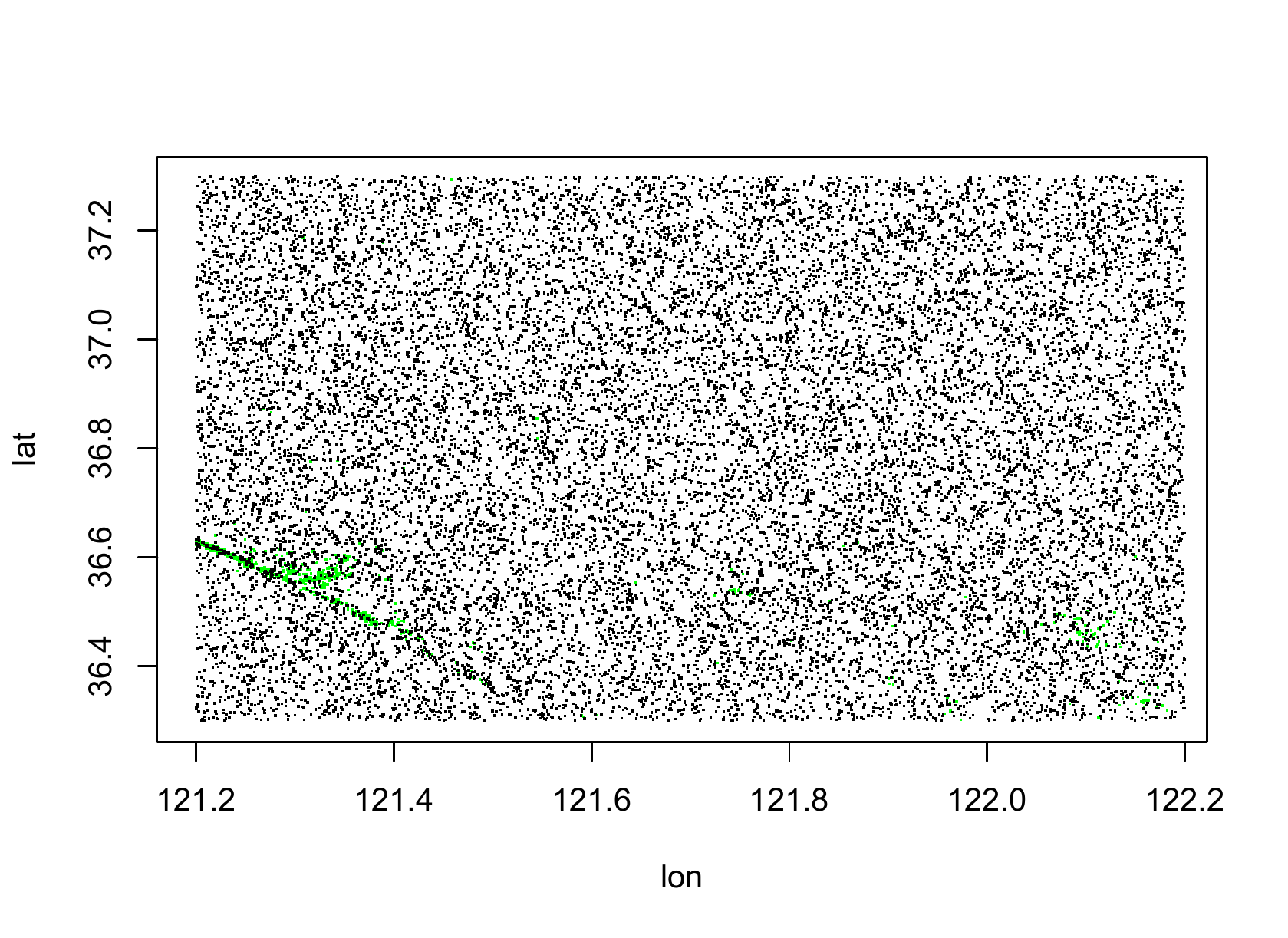}
\caption{Locations of super-thinned residuals using $b = $ 1 point/day after fitting variable productivity Hawkes model with productivity estimated using the truncated, rescaled and smoothed empirical estimate (black). Locations of original points corresponding to Bear Valley earthquakes are shown in green.} 
\label{fig:green} 
\end{figure} 

\begin{figure}[h] 
\includegraphics[height=3in,width=6in]{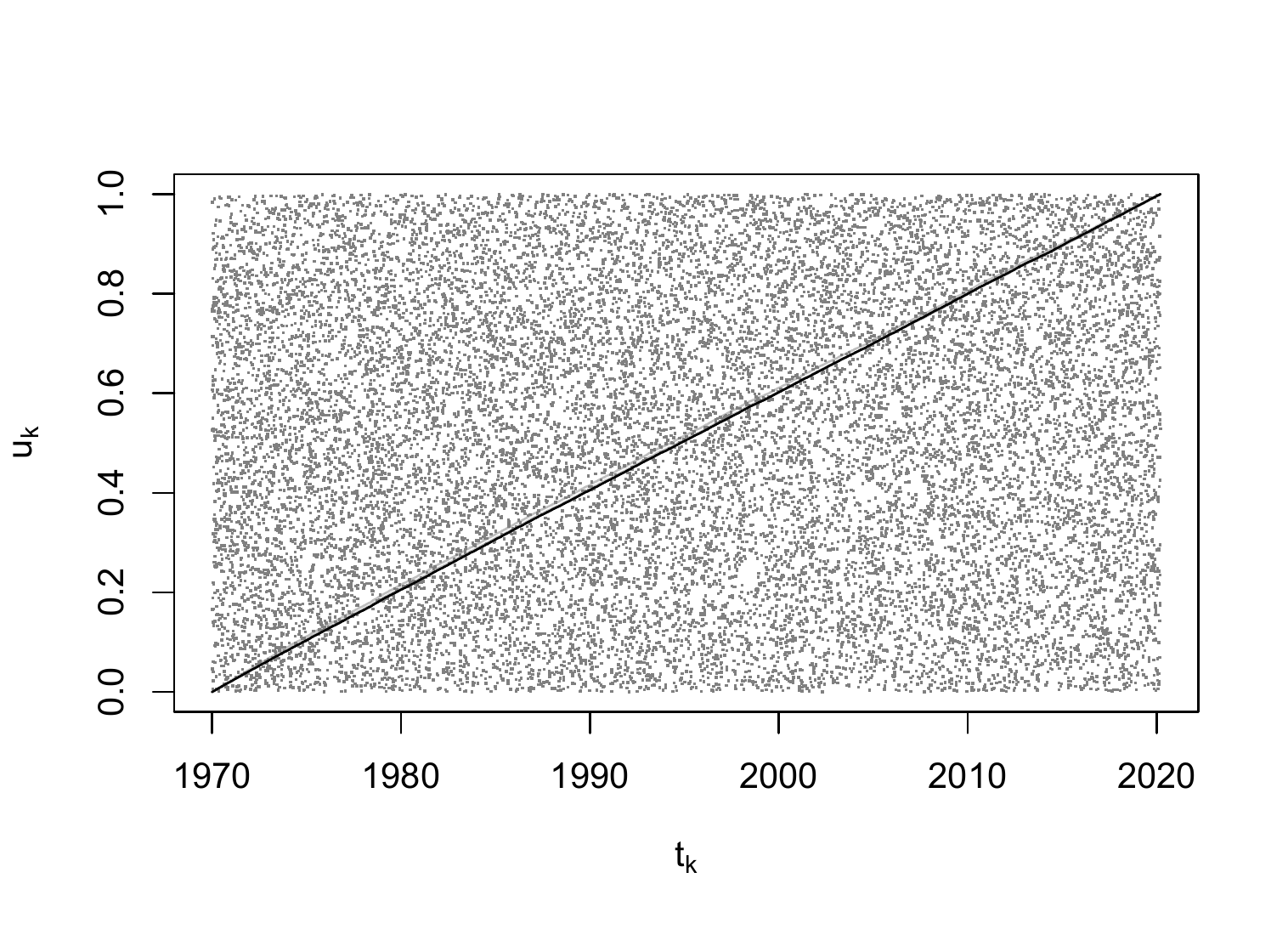}
\caption{Times $t_k$ of super-thinned residuals using $b = $ 1 point/day and their corresponding standardized interevent times $u_k$. The solid line shows, for each value of $t_k$, the normalized cumulative sum $\sum \limits _{i=1}^{k} u_i / \sum \limits _{i=1}^m u_i$, where $m$ is the number of super-thinned residuals. Dotted lines show lower and upper simultaneous 95\% confidence bounds based on 1000 simulations of the normalized cumulative sums of $m$ uniform random variables.} 
\label{fig:tkuk} 
\end{figure} 

The fit of the variable-productivity Hawkes model is assessed using super-thinned residuals (Clements et al.\ 2012). In super-thinning, one selects a constant $b$, thins the observations by keeping each observed point $\tau_i$ independently with probability $b/ \hat \lambda(\tau_i)$ if $\hat \lambda(\tau_i) > b$, 
and superposes points from a Poisson process with rate $(b - \hat \lambda) {\bf 1}_{\hat \lambda \leq b}$, where ${\bf 1}$ denotes the indicator function. 
The resulting super-thinned residuals form a homogeneous Poisson process with rate $b$ iff.\ $\hat \lambda$ is the true conditional rate of the observed point process (Clements et al.\ 2012). If $t_i$ are the times of the super-thinned points, then the interevent times, 
$r_i = t_{i} - t_{i-1}$, with the convention $t_0 = 0$, are exponential with mean $1/b$ if the fitted model $\hat \lambda$ is correct. One may thus inspect the uniformity of the standardized interevent times $u_i = F^{-1}(r_i)$, where $F$ is the cumulative distribution function of the exponential with mean $1/b$, as a means of goodness-of-fit assessment for the fitted model.\\ 

Figures \ref{fig:green} and \ref{fig:tkuk} shows the super-thinned residuals and their corresponding standardized interevent times $u_i$, as well as the cumulative sum of the standardized intervent times, along with individual 95$\%$ confidence bounds based on 1000 simulations of an equivalent number of uniform random variables. 
The super-thinned residuals appear to be approximately uniformly scattered in space and time, indicating no obvious lack of fit of the variable productivity Hawkes model.\\ 

\begin{figure}[h] 
\includegraphics[height=3in,width=6in]{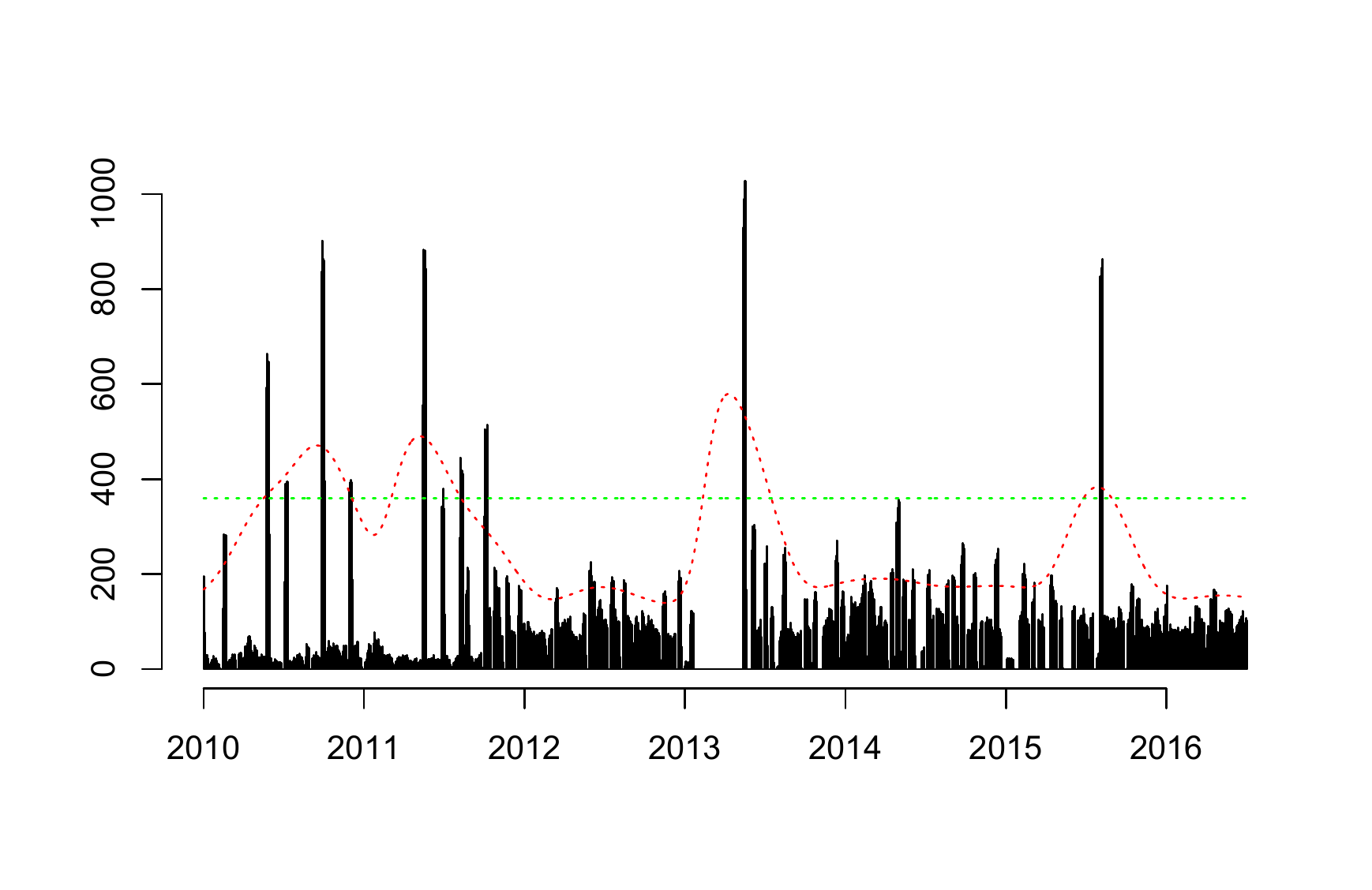}
\caption{Recorded numbers of Chlamydia cases in Arizona versus time, from 1/1/2010 to 7/2/2016, in cases/day (black), and estimated productivity according to Hawkes model (green) and variable productivity Hawkes model (red). Productivity estimates are in cases/yr. For Hawkes model and variable productivity Hawkes model, the triggering function $g$ is exponential with parameters $\beta$. For the Hawkes model, the parameters $\mu, K$, and $\beta$ are fit by MLE. The variable productivity Hawkes model uses the estimates of $\mu$ and $\beta$ fit by MLE for the Hawkes model, and the productivity is estimated using the truncated, rescaled, and smoothed empirical estimator smoothed over time. } 
\label{fig:histandprod} 
\end{figure} 

Hawkes models were also fit to recorded cases in Arizona from 1/1/2010 to 7/2/2016 of Chlamydial infection, a sexually transmitted disease. 
Cumulative counts during each year were obtained from Project Tycho (van Panhuis et al.\ 2018), www.tycho.pitt.edu , where the data were compiled from the United States Nationally Notifiable Disease Surveillance System. 
The records consist of cumulative totals within various time periods, and these cases were distributed uniformly during each observation window as in e.g.\ Meyer et al.\ (2012), Althaus (2014), Chaffee (2017), Schoenberg et al.\ (2018), Harrigan et al.\ (2019), Schoenberg et al.\ (2019), and Park et al.\ (2020). 
A total of 190,938 cases were recorded in Arizona during these 6.51 years.\\ 


Figure \ref{fig:histandprod} shows a histogram of the Chlamydia data along with estimates of the productivity 
according to an ordinary Hawkes model or a variable productivity Hawkes model. 
The MLEs of the parameters governing the Hawkes model $\lambda(t) = \mu + K \sum \limits {t_i < t} \beta exp(-\beta t-t_i)$ were fit to just the first two years of data, from 1/1/2010 to 1/1/2012, to prevent overfitting, resulting in the estimates $\{\hat \mu, \hat K, \hat \beta\} = \{1.177 (0.150), 0.984 (0.00467), 6.65 (0.150)\}$ points per day, with standard errors in parentheses obtained from the square roots of the diagonal elements of the inverse Hessian of the loglikelihood (Ogata 1978). The estimated productivity in Figure \ref{fig:histandprod} appears to vary substantially, suggesting an alternative to the ordinary Hawkes model might be appropriate.\\ 

\begin{figure}[h] 
\includegraphics[height=3in,width=6in]{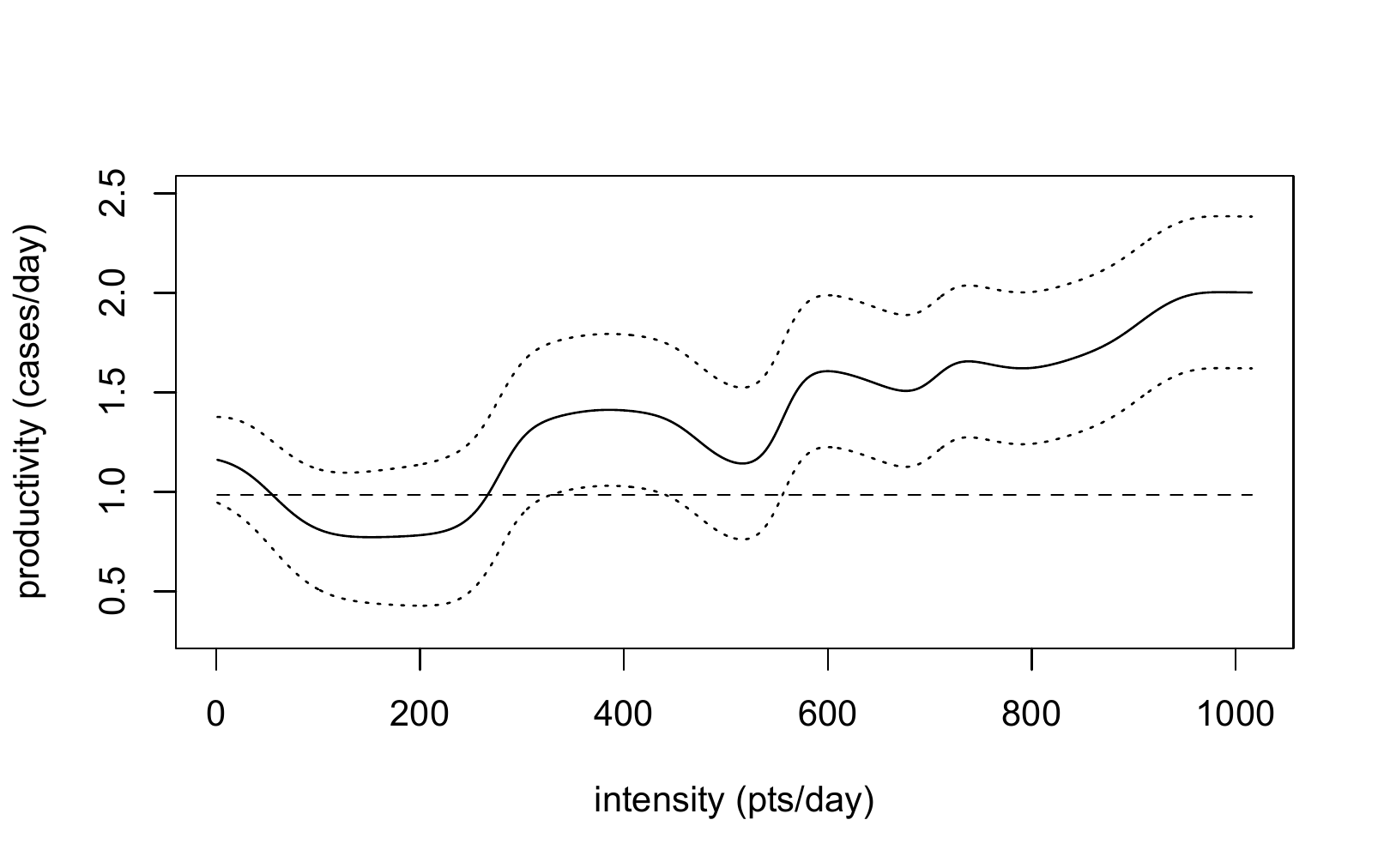}
\caption{Productivity estimates as functions of estimated intensity using the fitted Hawkes model 
$\hat \lambda(t) = \hat \mu + \hat K \sum \limits {t_i < t} \hat \beta exp(- \hat \beta t-t_i)$. Dashed curve indicates $\hat K$, the productivity estimated by MLE using the simple Hawkes model. Dotted curve indicates estimated productivity using the truncated, rescaled, and smoothed empirical estimator smoothed over $\hat \lambda$ using a normal kernel and bandwidth of 100. 
Dotted lines indicate estimates $\pm$ the standard deviation of estimates using simulations of the fitted simple Hawkes model and estimating the variable productivity using the truncated, rescaled, and smoothed empirical estimator smoothed over $\hat \lambda$ for each simulation. }
\label{fig:lamprodse} 
\end{figure} 

Figure \ref{fig:lamprodse} shows how the productivity estimates obtained using the truncated, rescaled, and smoothed empirical estimator vary as $\hat \lambda$ varies. 
There is an apparent though not statistically significant decrease in the estimated productivity as the estimated intensity increases from 0 to 200 points/day, using standard errors based on simulations. The fitted trend agrees with the hypothesis in Schoenberg et al.\ (2019) that productivities for infectious diseases may be lower when intensities are higher.  
$93.9\%$ of the times correspond to estimated intensities in this range. 
However, there is also a marked increase in estimated productivity at times of very high intensity. Possible explanations for this finding are explored in the next Section.\\ 

\section{Concluding remarks.}
 
The variable productivity Hawkes model fit to the Bear Valley earthquakes from 1970-2020 suggests significant departures from the exponential function of magnitude characteristic of productivity according to the ETAS model. 
It is possible, however, that the departures may be attributable to other sources of lack of fit in the ETAS model, such as inhonomogeneity in the background rate over this spatial temporal observation region. 
Although the superthinned residuals indicate the variable productivity Hawkes model appears to fit reasonably well, future research should examine whether the fit can be improved by allowing the background rate to be spatially inhomogeneous.\\ 

When applied to the 190,938 reported cases of Chlamydial infection in Arizona from 2010-2016, 
the fitted variable productivity Hawkes model revealed significant changes in productivity over time. 
The marked increase in estimated productivity associated with times of highest intensity for the Arizona Chlamydia data may be attributable to confounding factors, as it may be that certain environmental or economic factors or decisions by medical establishment may render Chlamydia more amenable to contagion or more likely to be reported at certain times rather than others, and thus may be associated with higher productivities and also higher intensities as a result. 
In particular, Chlamydia rates have been shown to vary substantially due to changes in screening, testing, and reporting procedures as well as variations in sexual behavior (Navarro et al.\ 2002, Shannon and Klausner 2018, 
and p25 of Centers for Disease Control and Prevention 2018).\\ 

In the case of earthquakes as well as Chlamydial infection, there may be numerous other covariates, such as climate, geographical and geological variables for instance, that are omitted here yet may influence the relationships observed describing the productivities of the points as a function of magnitude, time, or rate. 
Note however that the conditional intensity of the process may nevertheless be consistently estimated in the absence of such information provided the impact of the missing covariates is suitable small, as shown in Schoenberg (2016).\\ 

Future research should focus on improving the stability of the estimator (\ref{maineq}). 
To this end, we have shown that the main source of error in the estimates (\ref{maineq}) comes from instability in the estimation of the intensity using (\ref{lik2}). 
This suggests possibly improving the estimation of the productivities by improving the stability of the estimation of the intensities $\lambda(\tau_i)$ for $i=1,...,n$, 
perhaps by incorporating kernel intensity estimates along with equation (\ref{lik2}), or 
conditioning the matrix $G$ before computing its inverse, for example. 
More work is needed to adapt (\ref{maineq}) to the case where the adjacency matrix is singular or nearly singular, and to determine ideal means and bandwidths for smoothing the resulting estimates. 
In addition, future research should focus on whether the method proposed here can be extended to other types of point process models as well, such as Cox processes, inhibition processes, Gibbs point processes, or other models. 

\section*{Acknowledgements.}
This material is based upon work supported by the National Science Foundation
under grant number DMS 1513657. 
Waveform data, metadata, or data products for this study were accessed through the Northern California Earthquake Data Center (NCEDC), doi:10.7932/NCEDC . 
Computations were performed in R (https://www.r-project.org). 
Thanks to Project Tycho, the United States Nationally Notifiable Disease Surveillance System and the United States Centers for Disease Control and Prevention for supplying the data and for making it so easily available.\\ 

\pagebreak

\noindent {\sc \bf References}
\vskip 0.1in \parskip 1pt \parindent=1mm \def\reference{\hangindent=22pt\hangafter=1}

\reference



\reference 
Althaus, C.L. (2014). 
Estimating the reproduction number of Ebola virus (EBOV) during the 2014 outbreak in West Africa. 
{\sl PLOS Current Outbreaks} 6, doi: 10.1371/currents.outbreaks.91afb5e0f279e7f29e7056095255b288. 





\reference
Bacry, E., Mastromatteo, I., Muzy, J-F. (2015), 
'Hawkes processes in finance', 
{\sl Market Microstructure and Liquidity}, 1(1), 1550005.



\reference
Balderama, E., Schoenberg, F.P., Murray, E., and Rundel, P. W. (2012), 
'Application of branching models in the study of invasive species', 
{\sl Journal of the American Statistical Association}, 107(498), 467--476.








\reference 
Centers for Disease Control and Prevention (2018). 
{\sl Sexually Transmitted Disease Surveillance 2017.} 
U.S. Department of Health and Human Services, Centers for Disease Control and Prevention, 
National Center for HIV/AIDS, Viral Hepatitis, STD, and TB Prevention, Division of STD Prevention. 
Atlanta, Georgia, 168 pages. 

\reference 
Chaffee, A. (2017). 
Comparative Analysis of SEIR and Hawkes Models for the 2014 West Africa Ebola Outbreak. 
{\sl UCLA Electronic Theses and Dissertations}, University of California, Los Angeles, 1-33. 


\reference 
Clauset, A., and Woodard, R. (2013). 
Estimating the historical and future probabilities of large terrorist events. 
{\sl Ann.\ Appl.\ Stat.\ } 7(4), 1838--1865. 


\reference
Clements, R.A., Schoenberg, F.P., and Veen, A.,  2012. 
Evaluation of space-time point process models using super-thinning. 
{\sl Environmetrics} {\bf 23}(7), 606--616. 



\reference
Daley, D., and Vere-Jones, D. (2003), 
{\sl An Introduction to the Theory of Point Processes, Volume 1: Elementary Theory and Methods, 2nd ed.},
Springer: New York.

\reference
Daley, D., and Vere-Jones, D. (2007), 
{\sl An Introduction to the Theory of Point Processes, Volume 2: General Theory and Structure, 2nd ed.}
Springer: New York.

\reference 
Dascher-Cousineau, K., Brodsky, E.E., Lay, T., and Goebel, T.H.W.\ (2020). 
What controls variations in aftershock productivity? 
{\sl Journal of Geophysical Research: Solid Earth} 125, e2019JB018111, 1-18. 









\reference
Gordon, J. S., Clements, R. A., Schoenberg, F. P., and Schorlemmer, D. (2015),  
'Voronoi residuals and other residual analyses applied to CSEP earthquake forecasts', 
{\sl Spatial Statistics}, {\bf 14B}, 133--150. 




\reference 
Harrigan, R., M. Mossoko, E. Okitolonda-Wemakoy, F.P. Schoenberg, N. Hoff, P. Mbala, S.R. Wannier, S.D. Lee, S. Ahuka-Mundeke, T.B. Smith, B. Selo, B. Njokolo, G. Rutherford, A.W. Rimoin, J.J.M. Tamfum, and J. Park (2019). 
Real-time predictions of the 2018-2019 Ebola virus disease outbreak in the Democratic Republic of Congo using Hawkes point process models. 
{\sl Epidemics} 28, 100354. 

\reference 
Harte, D.S.\ (2014). 
An ETAS model with varying productivity rates. 
{\sl Geophysical Journal International} 198(1), 270--284. 

\reference
Hawkes, A. G. (1971), 
'Point spectra of some mutually exciting point processes', 
{\sl J. Roy. Statist. Soc.}, {\bf B33}, 438-443.

\reference
Meyer, S., Elias, J., and H\"ohle, S. (2012). 
A space-time conditional intensity model for invasive meningococcal disease occurrence.  
{\sl Biometrics} 68(2), 607--616. 



\reference 
Mohler, G.O., Short, M.B., Brantingham, P.J., Schoenberg, F.P., and Tita, G.E. (2011). 
Self-exciting point process modeling of crime. 
{\sl JASA}, 106(493), 100-108. 



\reference 
Navarro, C., Jolly, A., Nair, R., and Chen, Y. (2002). 
Risk factors for genital chlamydial infection. 
{\sl Can J Infect Dis.} 13(3), 195--207. 

\reference 
NCEDC (2014), Northern California Earthquake Data Center. UC Berkeley Seismological Laboratory. 
Dataset. doi:10.7932/NCEDC . 

\reference
Ogata, Y. (1978), 
'The asymptotic behavious of maximum likelihood estimators for stationary point processes', 
{\sl Annals of the Institute of Statistical Mathematics}, 30, 243--261.

\reference
Ogata, Y. (1988), 
'Statistical models for earthquake occurrence and residual
analysis for point processes', 
{\sl J.\ Amer.\ Statist.\ Assoc.}, {\bf 83}, 9-27.

\reference
Ogata, Y. (1998), 
'Space-time point-process models for earthquake occurrences',
{\sl Ann.\ Inst.\ Statist.\ Math.}, {\bf 50}(2), 379-402.


\reference 
Park J., Chaffee A., Harrigan R., Krebs A., and Schoenberg F.P. (2020). 
A non-parametric Hawkes model of the spread of Ebola in West Africa. 
{\sl J. Appl. Stat.}, to appear. 





\reference
Rathbun, S.L., Cressie, N. (1994), 
'Asymptotic properties of estimators for the parameters of spatial inhomogeneous Poisson point processes', 
{\sl Adv.\ Appl.\ Probab.}, 26, 122--154.




\reference 
Schoenberg, F.P. (2016), 
'A note on the consistent estimation of spatial-temporal point process parameters', 
{\sl Statistica Sinica}, 26, 861--789.

\reference 
Schoenberg, F., and Bolt, B. (2000). 
Short-term exciting, long-term correcting models for earthquake catalogs . 
{\sl Bulletin of the Seismological Society of America}, 90(4), 849--858.

\reference 
Schoenberg, F.P., Gordon, J.S., and Harrigan, R. (2018). 
Analytic computation of nonparametric Marsan-Lenglin\'e estimates for Hawkes point processes. 
{\sl Journal of Nonparametric Statistics} 30(3), 742-757. 

\reference 
Schoenberg, F.P., Hoffmann, M., and Harrigan, R. (2019). 
A recursive point process model for infectious diseases. 
{\sl AISM} 71(5), 1271-1287. 





Shannon C.L., and Klausner, J.D. (2018). 
The growing epidemic of sexually transmitted infections in adolescents: a neglected 
population. 
{\sl Curr.\ Opin.\ Pediatr.\} 30(1), 137--143. 



\reference 
B.W. Silverman (1986). 
{\sl Density Estimation for Statistics and Data Analysis.} 
Chapman and Hall, London. 





\reference 
van Panhuis W.G., Cross A., and Bruke D.S. (2018). 
Project Tycho 2.0: a repository to improve the integration and reuse of data for global population health. 
{\sl J. Am. Med. Inform. Assoc.} 25(12):1608?1617. 


\reference
Veen, A. and Schoenberg, F.P. (2008), 
'Estimation of space-time branching process models in seismology using an EM-type algorithm', 
{\sl. J. Amer. Statist. Assoc.}, 103(482), 614--624.





\reference 
Wetzler, N., Brodsky, E.E.\, and Lay, T. (2016). 
Regional and stress drop effects on aftershock productivity of large megathrust earthquakes. 
{\sl Geophysical Research Letters} 12,012-12020. 







\reference
Zechar, J. D., Schorlemmer, D., Werner, M.J., Gerstenberger, M.C., Rhoades, D.A., and Jordan, T.H. (2013), 
'Regional Earthquake Likelihood Models I: First-order results', 
{\sl Bull.\ Seismol.\ Soc.\ Am.\ }, 103, 787--798.



\end{document}